# Mapping spatial and temporal changes of global corporate research and development activities by conducting a bibliometric analysis

György Csomós
Department of Civil Engineering, University of Debrecen
E-mail: csomos@eng.unideb.hu

**Abstract**

Corporate research and development (R&D) activities have long been highly concentrated in a handful of world cities. This is due to the fact that these cities (e.g., Tokyo, New York, London, and Paris) are home to the largest and most powerful transnational corporations and are globally important sites for innovative start-up firms that operate in the fastest growing industries. However, in tandem with the rapid technological changes of our age, corporate R&D activities have shifted towards newly emerging and now globally significant R&D centres, like San Jose, San Francisco, and Boston in the United States, and Beijing, Seoul, and Shenzhen in East Asia. In this paper, I will conduct a bibliometric analysis to define which cities are centres of corporate R&D activities, how different industries influence their performance, and what spatial tendencies characterise the period from 1980 to 2014. The bibliometric analysis is based upon an assumption that implies there is a close connection between the number of scientific articles published by a given firm and the volume of its R&D activity.

    Results show that firms headquartered in Tokyo, New York, London, and Paris published the largest combined number of scientific articles in the period from 1980 to 2014, but that the growth rate of the annual output of scientific articles was much greater in Boston, San Jose, Beijing, and Seoul, as well as some Taiwanese cities. Furthermore, it can also be seen that those cities that have the largest number of articles; i.e., that can be considered as the most significant sites of corporate R&D in which firms operate in fast-growing industries, are primarily in the pharmaceutical and information technology industries. For these reasons, some mid-sized cities that are home to globally significant pharmaceutical or information technology firms are also top corporate R&D hubs.

# 1. Introduction

Since the beginning of the 1970s, but even in the past quarter century, the rate at which the global economy is restructuring has rapidly increased. This restructuring has been characterised by, for example, the emergence of the new international division of labour (NIDL) (Fröbel et al. 1980, Cohen 1981), as well as rapid technological changes (Dicken, 2007). Transnational corporations, as key orchestrators within the framework of economic globalization, have been relocating manufacturing activities away from core industrial countries towards developing countries for decades (Schoenberger, 1988; Dicken, 2007). As a result, some developing countries, especially China and India, have gradually become key actors in the world economy.

Recently in these developing countries, many giant low-tech manufacturing corporations and banks have been established, as well as a growing number of innovative small and medium-sized enterprises (Cheng, 2014), all of which have the power to significantly affect the world economy (Sauvant, 2008). Furthermore, in the recent past, technological change has occurred at a rapid pace due in part to fast-growing industries like nanotechnology, biotechnology, and information technology Hullmann, Meyer 2003; Nicolini, Nozza 2008; Dernis et al. 2015). Naturally, core economies still dominate in these industries; however, some developing countries have become serious competitors. For example, China has become a global player in the telecommunication and renewable energy industries. The gap between the economic performance of developed and developing countries has gradually narrowed, thanks not only to the fast economic growth of developing countries, but also to the large volume of foreign direct investments that target their corresponding R&D sector (Reddy, 2005). Several firms from developing countries have emerged as globally significant R&D investors (Hernández et al., 2015), and have themselves become major actors in global corporate R&D activities (Csomós, Tóth, 2016a).

Clearly, the spatial distribution of corporate R&D activities is highly uneven, because of the fact that the majority of R&D-oriented firms, and even their research facilities, are concentrated in just a few cities in the world. According to Sassen's (2001) global city concept, New York, London, Tokyo, and Paris are the most important sites for the production of innovation; i.e., they are expected to be the greatest centres of global corporate R&D activities. However, in tandem with the rapid technological changes of our age, corporate R&D activities have shifted towards newly emerging and globally significant R&D centres, like San Jose, San Francisco, and Boston in the United States (Rothwell et al., 2013), and Beijing, Seoul, and Shenzhen in East Asia. In recent years, these cities have become home to many R&D-oriented firms that did not even exist in the 1980s or 1990s.

In this paper, I will conduct a bibliometric analysis to define which cities are centres of corporate R&D activities, how different industries influence their performance, and what spatial tendencies characterise the period from 1980 to 2014. The structure of this paper is as follows: First, I will present the reasons firms have scientific articles; second, I will demonstrate the data and methodology by way of bibliometric analysis; and finally, I will draw the conclusion.

## 2. Research background

### 2.1. How can corporate R&D activities be measured?

The volume of corporate R&D activities can be measured in different ways; for example, by counting the number of patents and/or the number of patent citations (Narin et al., 1987; Chang et al., 2012; Liu et al., 2006; Ribeiro et al., 2010; Ribeiro et al, 2014; Wang et al., 2011; Wong, Wang, 2015); by counting the amount in R&D expenditures (Granstrand, 1999; Kumar, 2001; Piergiovanni, Santarelli, 2013; Yoo, Moon, 2006); by defining the quantity and quality of research cooperation between firms and universities (Feng et al., 2015; Gao et al., 2011; Kneller et al., 2014; Ramos-Vielba et al., 2010); and by the number of scientific articles authored or co-authored by researchers within the firms (Chang, 2014; Furukawa, Goto, 2006; Hicks et al., 1994; Hicks, 1995; Hullmann, Meyer, 2003; Tijssen, 2004; Csomós, Tóth, 2016a). The latter method is confirmed by Narin et al. (1987: 144), who addressed the product life cycle of corporations:

*"Corporate sales lead to corporate profits, which may be used in research and development to produce scientific innovations (scientific publications), which may lead to technological innovations (some of which will be patented), which is the catalyst for new products and more efficient processes, which will increase corporate profits."*

Thus, the number of scientific articles authored/co-authored by corporate researchers or engineers can reflect upon the R&D capacity of firms.

## 2.2. Why do firms have scientific publications?

Not every firm publishes scientific papers; not even some that have many patents or a large budget for R&D expenditures. Although, it is generally believed that scientific publishing is out of business organizations' interest, firms publish papers, and they do it extensively (Hicks, 1995; Godin, 1996). Some firms contribute as many scientific articles to public literature as do medium-sized universities. For example, between 1980 and 2014, IBM published more scientific articles than did Carnegie Mellon University, which was ranked 61$^{st}$ on the 2015 ARWU ranking. Firms' motivations regarding publishing differs between countries and across industries, and is largely influenced by the firm's individual intellectual property strategy.

Next, I will provide an overview on the main reasons that firms contribute scientific articles, or at minimum, why firms have scientific articles indexed by bibliographic databases (e.g., Web of Science, Scopus).

### 2.2.1. Enabling defensive publications

A firm that has a patentable innovation can choose from any or all of these three options. First, the firm can obtain a patent; second, it can maintain trade secrecy; and third, it can publish an enabling defensive publication[1] (Johnson, 2014). Each option has benefits and risks that must be considered by the firm. The primary cost of a defensive publication is that it discloses the technical information of a product or process to the public, allowing competitors to capitalise on the innovation free of charge. Consequently, the innovating firm loses the right to exclude others from producing, selling, or using the patented innovation for the term of the patent, which is usually 20 (+5 extra) years from the filing date. The main benefit of an enabling publication is that it destroys patent rights. According to Barrett (2002: 191), "the successful defensive publication renders the competitor's invention obvious or lacking in novelty."

Enabling defensive publications has been a component of comprehensive intellectual property (IP) strategies of firms, even the largest transnational corporations, for decades. IBM produced its Technical Disclosure Bulletin between 1958 and 1998 to publish defensive disclosures of inventions that were not formally patented. By having done this, IBM lost the possibility of patenting a number of innovations; however, the cumulative benefits enjoyed by the firm were much higher by destroying competitors' patent rights (Barrett, 2002). The Bulletin, as prior art, is referred to in a US patent document more than 48,000 times. Recently, defensive publication has become a widely used tool for small and medium-sized enterprises (SMEs) to protect intellectual property instead of patenting innovations. This choice of strategy has several reasons behind it; however, the two most important ones are that patent applications are too expensive for SMEs (primarily in the United States), and it takes an average of two years from the date of filing to process an application (The New York Times, 2002).

---

[1] There are three options to protect innovations and inventions (www.defensivepublications.org):
1) Obtain a patent. A patent is the right, granted by a government (in exchange for full disclosure), to exclude others from making, using or selling products and services for a period of time (generally twenty years from the filing date).
2) Maintain secrecy. Innovations or inventions that are safeguarded from disclosure to others by policies and procedures are trade secrets. They are generally protected as long as secrecy is maintained.
3) Publish an enabling defensive publication.

The phenomenon of defensive publication significantly contributes to the increase in the number of firms' scientific publications[2].

### 2.2.2. Enhancing the reputation of the firm

Many firms encourage their researchers and engineers to disclose knowledge publicly in order to improve the external reputation of the firm (Hicks, 1995; Li et al., 2015). The improved reputation achieved via open knowledge disclosure can result in many benefits for a given firm (Muller, Pénin, 2006). First, reputable firms can more easily find investors or obtain grants and subsidies. Second, they can more easily find potential collaborators with whom they can work on joint R&D projects. Third, these firms are more readily accepted for involvement in broader academic and industrial networks, resulting in additional benefits for them. Fourth, these firms can attract and hire star researchers to work on the firms' R&D projects. The improved reputation of the firm may counterweight the losses in profits (Allen, 1983), furthermore, reputational gains from being published in scientific publications give credibility to a firm in its research field (Li et al., 2015).

It can be concluded that scientific publication by firms can be considered on some level as quasi-scientific advertisements to target other innovative firms and researchers, informing them of the firm's leadership in relevant R&D projects.

### 2.2.3. Recruiting and retaining researchers

According to Godin (1996), firms are not generally interested in publishing scientific papers because their main motivation is to develop patentable innovations, which increases corporate profits. Therefore, before patenting innovations, researchers are not permitted by their firm to publish their results (i.e., disclose the technical information behind innovations to the public). The reasoning is apparent: Patents are valuable in terms of money, while scientific publications are not (unless defensive publishing is a component of the firm's IP strategy) (Levi-Mazloum, von Ungern-Sternberg, 1990; Pain, 2009). This restriction against publishing may discourage many researchers from working for firms for a higher salary but a lower scientific reputation, as many are more likely to be committed to working in academia for a lower salary but a greater scientific reputation. This conflict may lead to an extreme situation: either firms not being able to hire researchers for their R&D projects, or researchers leaving the firm upon being restricted from publishing. In order to avoid this situation, a compromise must be made between the management and researchers, in which researchers ask permission before publishing their scientific results, with cooperation from the management in allowing researchers to publish (Furukawa & Goto, 2006). This compromise has key benefits for both parties.

---

[2] Defensive publications do not necessarily correspond to scientific publications. The Ip.com Prior Art Database contains the largest number of enabling defensive publications in the world. Although these defensive publications may provide new knowledge by disclosing technical information about innovations, they are not considered scientific publications because they are not, among other things, peer-reviewed, not indexed in citation databases, etc.

**2.2.4. Collaboration with universities**

Collaboration between firms and universities on R&D projects has always had great significance. According to Kneller et al. (2014), "motivations for firms to engage with universities include accessing complementary research expertise for future business development, particularly for products that are in the design or early development stage, furthermore, small or new firms tend to rely on universities for their core technologies." The results of this kind of collaboration can be patentable innovations (which are important to the firm), as well as joint scientific publications (which are important to academicians) (Ramos-Vielba 2010). In many cases, corporate researchers and engineers contribute to the writing of a paper only by providing technical guidance, professional supervision, and necessary data; they do not participate directly in the creation of the paper. However, as quasi co-authors, their names and their affiliations appear in the paper, which then will be indexed by bibliographic databases. Thus, the firm, as the affiliate of corporate researchers, will enjoy credit in a scientific publication, even if it was not directly involved in writing the paper.

**3. Data collection**

In this paper, I conducted a bibliometric analysis of data in the Scopus database (www.scopus.com). Scopus is the largest abstract and citation database, containing almost 22,000 titles (20,800 peer-reviewed journals, 367 trade publications, and more than 400 book series) from 5,000 publishers, in addition to 6.4 million conference papers. Scopus offers the most broad and integrated coverage available of scientific, technical, medical, and social sciences including arts and humanities literature. In this analysis, I focus on full articles exclusively, leaving out of consideration all other types of publications (for example, conference papers, book chapters, letters, and editorials), which do not necessarily go through a peer-review process. Articles, written by given firms' researchers and engineers, have been assigned to cities as was indicated by Scopus.

The 2015 EU Industrial R&D Investment Scoreboard ranks the world's top 2,500 R&D-investing firms, which together invested $681.9 billion in R&D, representing about 90 percent of the world's total expenditure by businesses on R&D. Leading R&D investor firms are classified into 41 industry sectors, and are headquartered in 475 cities across 40 countries worldwide (Hernández et al., 2015). However, only 1,027 out of the 2,500 firms have scientific articles indexed in Scopus, and these firms are headquartered in 261 cities. Researchers and engineers at these firms contributed a total number of 958,725 scientific articles in the period from 1980 to 2014. Furthermore, it is needed to be mentioned that articles can be written by

multiple authors who affiliate to multiple organizations, i.e. a given article can be assigned to not only one, but multiple corporations.

## 4. Results

### 4.1. General results of corporate scientific publishing

#### 4.1.1. Correlation between industry sectors and number of scientific articles

Previous studies underline the fact that most R&D-oriented firms are involved in high-tech industries (e.g., pharmaceuticals, information technology, chemicals, and electronics), and as such, they publish the largest number of scientific publications (Godin, 1996; Chang, 2014). Table 1 shows that the Pharmaceuticals & Biotechnology industry contributed nearly 29 percent of the total number of articles, even though the share of the number of firms that operate in this industry is under 13 percent. Thanks to Nippon Telegraph & Telephone (NTT), the number of articles per firm is actually the largest in the Fixed Line Telecommunications industry, in which NTT contributed 64 percent of the total number of articles.

**Table 1. Ranking industry sectors by the number of scientific articles published by firms in the period from 1980 to 2014**

| Rank | Industrial sector (ICB-3D) | No. of firms | No. of articles (1980-2014) | No. of articles per firm | Percentage within the dataset |
|---|---|---|---|---|---|
| 1 | Pharmaceuticals & Biotechnology | 317 | 276,215 | 871 | 28.81 |
| 2 | Technology Hardware & Equipment | 317 | 103,874 | 328 | 10.83 |
| 3 | Electronic & Electrical Equipment | 229 | 79,406 | 347 | 8.28 |
| 4 | General Industrials | 96 | 75,002 | 781 | 7.82 |
| 5 | Software & Computer Services | 275 | 69,606 | 253 | 7.26 |
| 6 | Chemicals | 133 | 62,012 | 466 | 6.47 |
| 7 | Oil & Gas Producers | 32 | 53,771 | 1680 | 5.61 |
| 8 | Automobiles & Parts | 155 | 45,379 | 293 | 4.73 |
| 9 | Fixed Line Telecommunications | 17 | 35,267 | 2075 | 3.68 |
| 10 | Aerospace & Defence | 56 | 24,233 | 433 | 2.53 |
| 11 | Industrial Metals & Mining | 40 | 18,710 | 468 | 1.95 |
| 12 | Industrial Engineering | 199 | 17,795 | 89 | 1.86 |
| 13 | Food Producers | 59 | 17,809 | 302 | 1.86 |
| 14 | Leisure Goods | 39 | 16,247 | 417 | 1.69 |
| 15 | Health Care Equipment & Services | 100 | 13,132 | 131 | 1.37 |
| 16 | Electricity | 30 | 10,799 | 360 | 1.13 |
| 17-41 | Other industry sectors | 412 | 39,468 | - | 4.12 |
| | TOTAL | 2506 | 958,725 | 383 | 100.00 |

Table 2 shows those industry sectors of cities in which more than 10,000 articles were created in the period from 1980 to 2014. Recognizing the fact that the largest number of articles belongs to pharmaceuticals, it is not surprising that the Pharmaceuticals & Biotechnology category is the most important industrial component of leading cities. For example, 23 pharmaceutical firms headquartered in New York published as many articles as did 275 information technology firms (classified into the Software & Computer Services category) in the period from 1980 to 2014. Furthermore, some leading cities (Tokyo, New York, and Paris) had more than one industry in which firms published a very large number of articles; i.e., they carried out extensive R&D activities.

**Table 2. Ranking industry sectors in cities by the number of scientific articles published by firms**

| Rank | City/Metro | Country | Industrial sector (ICB-3D) | No. of firms | No. of articles (1980-2014) | Percentage within the sector |
|---|---|---|---|---|---|---|
| 1 | New York | United States | Pharmaceuticals & Biotechnology | 23 | 69,871 | 25.30 |
| 2 | London | United Kingdom | Pharmaceuticals & Biotechnology | 8 | 42,646 | 15.44 |
| 3 | New York | United States | Software & Computer Services | 13 | 39,742 | 57.10 |
| 4 | Basel | Switzerland | Pharmaceuticals & Biotechnology | 4 | 39,062 | 14.14 |
| 5 | Tokyo | Japan | Electronic & Electrical Equipment | 24 | 35,548 | 44.77 |
| 6 | Paris | France | Technology Hardware & Equipment | 4 | 32,786 | 31.56 |
| 7 | New York | United States | General Industrials | 4 | 26,172 | 34.90 |
| 8 | San Jose | United States | Technology Hardware & Equipment | 63 | 22,502 | 21.66 |
| 9 | Tokyo | Japan | Fixed Line Telecommunications | 1 | 22,467 | 63.71 |
| 10 | Beijing | China | Oil & Gas Producers | 3 | 18,575 | 34.54 |
| 11 | Paris | France | Pharmaceuticals & Biotechnology | 8 | 17,885 | 6.48 |
| 12 | Tokyo | Japan | Software & Computer Services | 6 | 17,126 | 24.60 |
| 13 | Tokyo | Japan | General Industrials | 10 | 16,670 | 22.23 |
| 14 | Tokyo | Japan | Pharmaceuticals & Biotechnology | 15 | 16,654 | 6.03 |
| 15 | Amsterdam | Netherlands | General Industrials | 1 | 14,385 | 19.18 |
| 16 | Cologne | Germany | Pharmaceuticals & Biotechnology | 1 | 13,538 | 4.90 |
| 17 | Munich | Germany | Electronic & Electrical Equipment | 2 | 13,252 | 16.69 |
| 18 | Tokyo | Japan | Automobiles & Parts | 21 | 13,022 | 28.70 |
| 19 | Indianapolis | United States | Pharmaceuticals & Biotechnology | 1 | 12,765 | 4.62 |
| 20 | Philadelphia | United States | Chemicals | 3 | 11,972 | 19.31 |
| 21 | Tokyo | Japan | Chemicals | 28 | 10,862 | 17.52 |
| 22 | Detroit | United States | Automobiles & Parts | 10 | 10,678 | 23.53 |
| 23 | Nagoya | Japan | Automobiles & Parts | 12 | 10,139 | 22.34 |
| 24 | Bridgeport | United States | General Industrials | 1 | 10,031 | 13.37 |
| 25-602 | Cities' other industry sectors | | | 1410 | 420,375 | |
| | TOTAL | | | 1676 | 958,725 | |

**4.1.2. Cities as centres of global corporate R&D**

The preceding industry-specific ranking suggests that Tokyo and New York are leading sites of corporate R&D. Table 3 illustrates that these cities were home to two-thirds of the total number of articles published by selected firms in the period from 1980 to 2014. Tokyo is the leading headquarter location of corporate R&D investors, as headquarters to nine percent of all firms in the data set. As home to 78 firms on the list, New York is also a significant headquarters city; moreover, New York-based IBM, Pfizer, Honeywell, and Merck are all Top 10 firms in terms of the number of scientific articles published. It is not surprising that Paris and London, two European global cities, are in leading positions, because both cities are strong pharmaceutical industry bases (London-based firms include GlaxoSmithKline and AstraZeneca; Paris-based firms include Sanofi and Servier), furthermore Paris is home to Alcatel-Lucent, which published the second-largest number of scientific articles in the 1980 to 2014 time frame, after IBM.

**Table 3. Ranking cities by the total number of scientific articles published by headquartered firms**

| Rank | City/Metro | Country | No. of firms | No. of articles (1980-2014) | Percentage within the dataset |
|---|---|---|---|---|---|
| 1 | Tokyo | Japan | 227 | 171,917 | 17.93 |
| 2 | New York | United States | 78 | 140,329 | 14.64 |
| 3 | Paris | France | 70 | 72,173 | 7.53 |
| 4 | London | United Kingdom | 84 | 60,751 | 6.34 |
| 5 | Basel | Switzerland | 8 | 41,098 | 4.29 |
| 6 | San Jose | United States | 110 | 29,088 | 3.03 |
| 7 | Osaka | Japan | 78 | 26,774 | 2.79 |
| 8 | Chicago | United States | 34 | 20,220 | 2.11 |
| 9 | Beijing | China | 63 | 19,607 | 2.05 |
| 10 | Amsterdam | Netherlands | 14 | 19,545 | 2.04 |
| 11 | Seoul | South Korea | 66 | 18,629 | 1.94 |
| 12 | Munich | Germany | 22 | 17,127 | 1.79 |
| 13 | Dallas | United States | 12 | 16,879 | 1.76 |
| 14 | Bridgeport | United States | 9 | 16,244 | 1.69 |
| 15 | Boston | United States | 84 | 14,549 | 1.52 |
| 16 | Cologne | Germany | 4 | 13,656 | 1.42 |
| 17 | Philadelphia | United States | 23 | 12,960 | 1.35 |
| 18 | Indianapolis | United States | 3 | 12,765 | 1.33 |
| 19 | Nagoya | Japan | 32 | 11,876 | 1.24 |
| 20 | Detroit | United States | 11 | 10,678 | 1.11 |
| 21 | Washington | United States | 24 | 10,407 | 1.09 |
| 22 | Oxnard | United States | 4 | 7919 | 0.83 |
| 23 | San Francisco | United States | 89 | 7292 | 0.76 |
| 24 | Houston | United States | 15 | 7069 | 0.74 |
| 25 | The Hague | Netherlands | 2 | 6981 | 0.73 |
| 26 | Copenhagen | Denmark | 16 | 6848 | 0.71 |
| 27 | Seattle | United States | 16 | 6591 | 0.69 |
| 28 | Zurich | Switzerland | 25 | 6426 | 0.67 |
| 29 | Mainz | Germany | 1 | 5843 | 0.61 |
| 30 | Helsinki | Finland | 20 | 5531 | 0.58 |
| 31 | Minneapolis | United States | 17 | 5256 | 0.55 |
| 32 | Cincinnati | United States | 3 | 5098 | 0.53 |
| 33 | Midland, Michigan | United States | 1 | 5082 | 0.53 |
| 34 | Ludwigshafen | Germany | 1 | 5056 | 0.53 |
| 35-261 | Other cities | - | 929 | 120,461 | 12.56 |
|  | TOTAL |  | 2195 | 958,725 | 100.00 |

The four global cities, New York, London, Tokyo, and Paris, are followed by Basel (Switzerland) and San Jose (United States). Basel is a leading site in the international pharmaceutical industry: Basel-based Novartis and Roche both belong to Top 10 firms in terms of the number of published scientific articles. Table 3 shows that Basel is home to only eight significant corporate R&D investors, but San Jose is the headquarters of 110 relevant firms. In the period from 1980 to 1984, the listed Basel-based firms published 2.4 times as many articles as did the firms in San Jose. This ratio lessened to 1.2:1 in the period from 2009 to 2014, emphasizing that the gap between Basel and San Jose is closing.

It should be noted that in the period from 1980 to 2014, the vast majority of the total number of scientific articles (87.44 percent) were published by firms located in developed countries. Out of the 34 top-ranked cities (see Table 3), 17 cities are located in the United States, 12 in Western Europe, three in Japan, and one in South Korea. The category of leading cities in developing countries is represented only by Beijing.

**4.2. The most intensive period of publishing**

The United States is the dominant actor in corporate R&D activities. In the period from 1980 to 2014, 38 percent of the total number of scientific articles came from US firms. Western Europe and Japan are both important locations of corporate R&D activities as well, publishing 33 percent and 22 percent of the total number of articles, respectively. All the other countries play a minor role in global corporate R&D, contributing only seven percent of the total number of articles. It is important to now analyse the spatial dynamics of scientific publishing.

Figure 1 shows the most intensive period of publishing by city; i.e., a five-year-long period in which firms located in a given city published the largest number of scientific articles. Cities are classified into three groups, depending on whether their most intensive period of publishing fell into the time frame 1) between 1980 and1984, and 1999 and 2003; 2) between 2000 and 2004, and 2009 and 2013; or 3) between 2010 and 2014. It can be seen that the three major areas: the United States, Western Europe, and Japan, are spatially divided regarding their most intensive period of publishing. Perhaps Japan is an exception, because its key cities' most intensive period came to an end before 2013; moreover, in the cases of Tokyo and Osaka, the two most important cities in terms of the number of articles, this five-year-long period fell in the 1990s.

Many major cities on the West Coast and the East Coast of the United States have become international centres of corporate R&D activities, while cities in the Great Lakes region and the state of Texas have lost their leading positions. Information technology firms have contributed the largest number of scientific articles in Seattle and San Jose, while in the cases of Oxnard, San Diego, and San Francisco, pharmaceutical firms have published a great number of articles. On the East Coast of the United States, Boston-based biotechnology and pharmaceutical firms have contributed the largest number of articles; furthermore, some highly specialised firms located in small and medium-sized cities have also published a great number of articles (e.g., Corning, New York: Corning Inc.; Hartford, Connecticut: United Technologies).

The Great Lakes region and the state of Texas are home to many transnational corporations that operate in traditional industries, like oil, chemicals, and automotive. Some of these firms had once been among the world's largest firms (e.g. General Motors and ExxonMobil), but after the recession of the late 2000s, and in part due to the recent low oil prices, they have come into a crisis, or in some cases, have been recovering slowly. Because of their financial problems, these firms have cut their R&D expenditures and reduced the size of the R&D staff, which has naturally resulted in a decrease in the number of scientific articles.

In the period from 1980 to 2014, the largest number of scientific articles in the United States was published by New York-based firms. New York is home to several global pharmaceutical firms, including Merck, Pfizer, Johnson & Johnson, and Bristol-Myers Squibb, whose most intensive year of publishing was 2011; however, since that time, there has been a significant decrease. The peak of scientific publishing within the information technology industry, which is the second most dominant industry sector in New York in terms of scientific output, was 1989 to 1993, but since that time, it has also witnessed a decrease in the number of articles published.

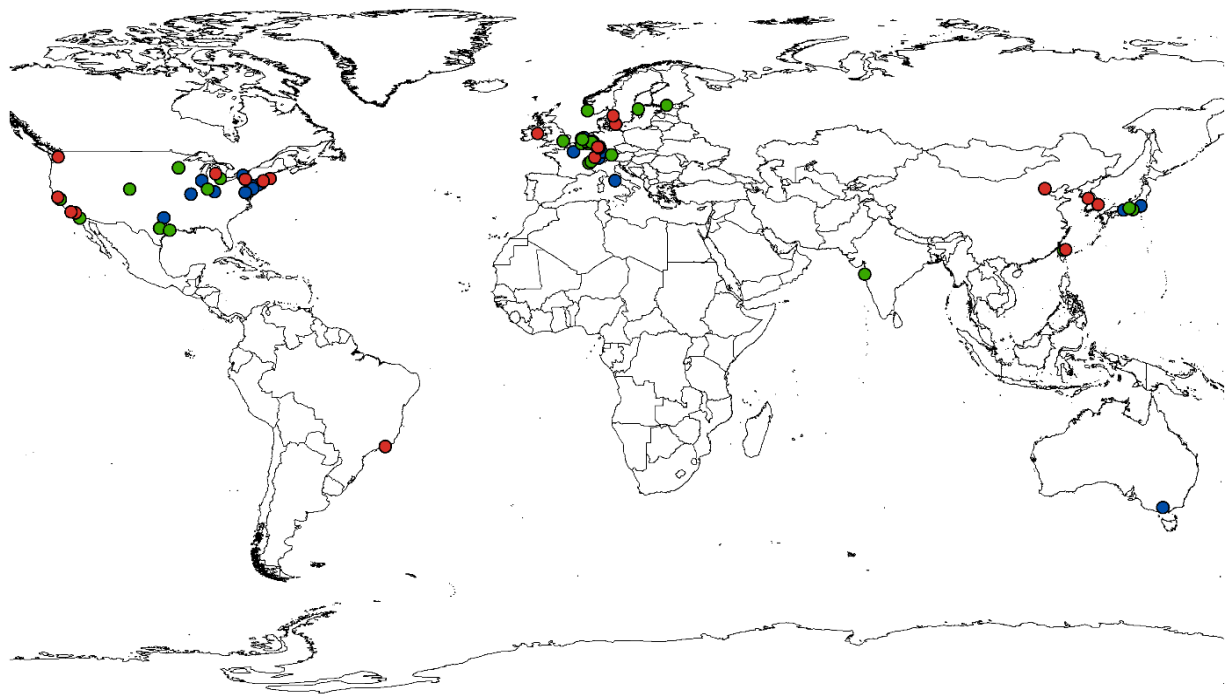

**Fig. 1. Mapping spatial distribution of leading cities as centres of corporate R&D in light of the most intensive period of publishing**

In Europe, cities that are home to large pharmaceutical firms; for example, Dublin, Copenhagen, Basel, and Darmstadt, showed a marked increase in the number of scientific articles. The most surprising among these cities is Dublin, a relative newcomer to the pharmaceutical industry. Basel and Darmstadt have long been home to major pharmaceutical firms: the histories of Basel-based Roche and Novartis, and that of Darmstadt-based Merck, are inseparable from the histories of their headquarter cities. Copenhagen is the European centre of several medium-sized pharmaceutical firms and fast-growing biotechnology firms, which were all founded in Denmark. In contrast with these cities, Dublin has attracted pharmaceutical firms (and some more large research-oriented firms) from other countries, especially from the United States. US firms, including Allergan, Perrigo, Eaton, Seagate, and Accenture, relocated their headquarters, and in some cases, the entire firm, to Dublin, because the Irish capital offered low corporate taxes and well-developed business infrastructure (The New York Times, 2015).

However, as illustrated in Table 3, the most significant European sites for corporate R&D were Paris and London. Firms in these two cities published the largest number of scientific articles in the period from 1980 to 2014. For London-based firms, the most intensive period of publishing was 2007 to 2011, but by 2010 to 2014, a decrease of less than five percent occurred. In contrast, firms in Paris published the largest number of articles in the period from 1996 to 2000, and since then the city showed a 46 percent decline by 2010 to 2014. In London, two global pharmaceutical firms, GlaxoSmithKline and AstraZeneca, published 59 percent of the articles. The decline in London's annual output was mainly attributed to GlaxoSmithKline, which decreased its publishing activity by nine percent by 2014. In Paris, there were two dominant firms as well: Alcatel-Lucent, a global telecommunications equipment firm, and Sanofi, one of the world's largest pharmaceutical firms. These two firms published 64 percent of all articles of Paris. Alcatel-Lucent contributed about 33,000 articles in the period from 1980 to 2014, about 45 percent of the total number of the Paris-based firms' scientific articles. Alcatel-Lucent published 7,323 articles in its most intensive period of 1987 to 1991. Its less intensive period was 2010 to 2014, when the firm only contributed 1,095 articles, which translates to an 85 percent decrease in the number of articles over the course of 15 years. Since the beginning of the 2000s, Alcatel has desperately attempted to compete with its Asian rivals (such as ZTE and Huawei), and even the company's 2006 acquisition of the American Lucent did not result in success (Financial Times, 2012). Between 2006 and 2015, the merged Alcatel-Lucent suffered a cumulative loss of $13 billion; meanwhile, both its revenues and market values fell significantly. In 2006, Alcatel also acquired Lucent's Bell Laboratories, one of the world's largest research facilities and owner of a great number of patents and scientific publications. Alcatel-Lucent still had a globally outstanding R&D expenditure in 2013, but it was not as much as the independent firms' combined expenditures had been in the 1990s (also reflected by the fact that the number of scientific articles has gradually decreased since the late 1980s). Alcatel-Lucent could not avoid selling its shares in other firms (e.g., in Thales and Genesys), or even its own acquisition. In 2016, Finland-based Nokia, one of the world's largest telecommunication network equipment firms, acquired Alcatel-Lucent and Bell Labs, and organised them into quasi-autonomous subsidiaries. In conclusion, up until 2014, Paris maintained a very good position in global corporate R&D, but after losing control of Alcatel-Lucent, its leading role is fading away.

Since the mid-1990s, significant changes have occurred in the spatial structure of corporate R&D in East Asia. The publishing activity by major Japanese firms that ruled the 1990s has been gradually decreasing. For example, Tokyo- and Osaka-based firms published the largest number of scientific articles in 1996 respectively, but their annual output decreased by 45 percent by the year 2014. Regarding corporate publishing activity, there is no dominant firm in Tokyo. The city's 10 largest firms in terms of the number of articles contributed only 61 percent to the total number of articles published by all Tokyo-based firms. Although the annual output of articles has been decreasing in almost every industry sector, the negative change far exceeds the average in the electronic industry, one of Japan's most important industries (The Economist, 2009). Among the Tokyo-based firms that contributed the largest number of articles were several electronic firms, including Hitachi (ranked 2[nd]), NEC (5[th]), Toshiba (6[th]), and Sony (11[th]), whose collective average annual output decreased by 34 percent from 1996 to 2014. In Osaka, another electronics giant, Panasonic, contributed the largest number of articles over the course of the study. Its most intensive year of publishing was in the mid-1990s; however, its annual output had decreased by 50 percent by the year 2014. Further adding to the losses, the fourth-largest Osaka-based firm in terms of the number of articles published, Sharp, has since been acquired by the Taiwanese Foxconn.

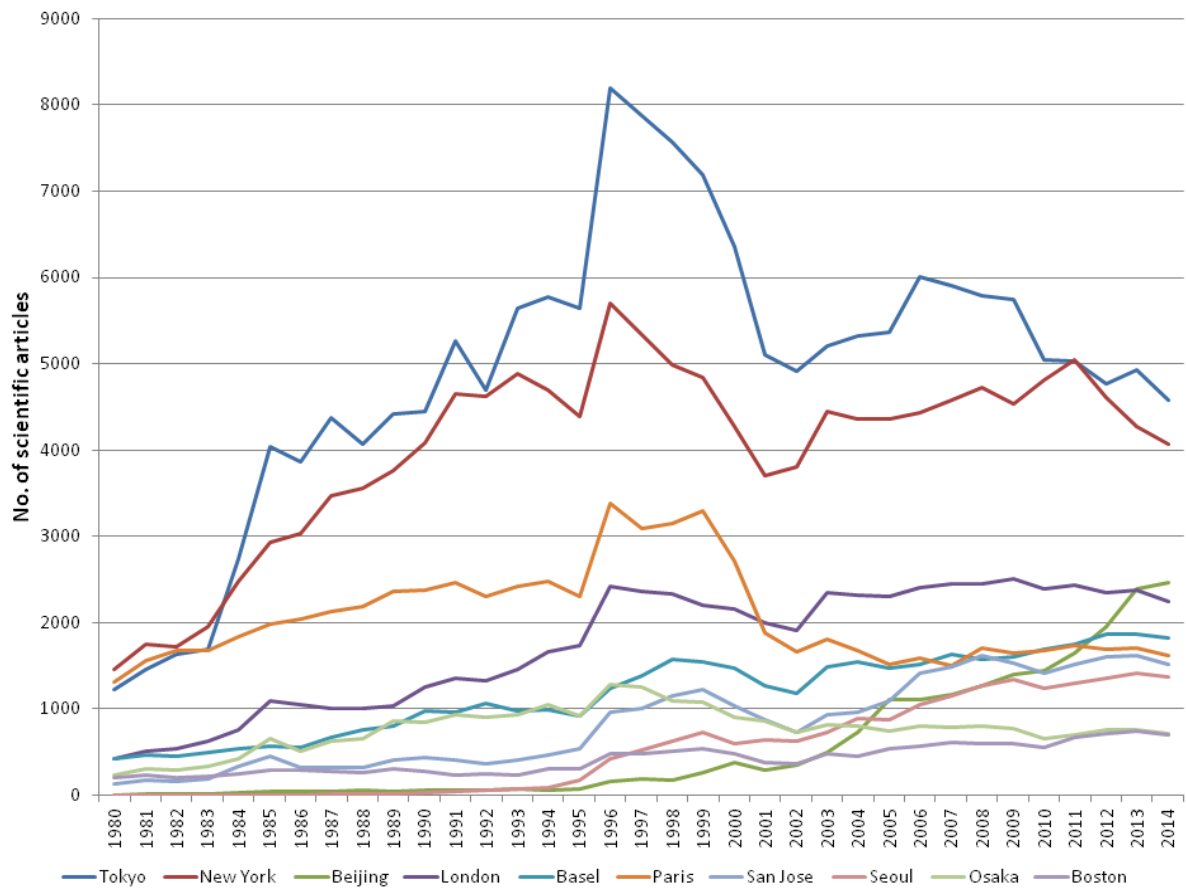

**Fig. 2. Annual output of scientific articles by leading firms in Top 10 cities (1980-2014)**

In East Asia, Seoul and Beijing have recently shown a great increase in the number of articles published. Today these cities have become major competitors to the Japanese cities, not only as command and control centres (see, for example, Csomós, Tóth, 2016b), but also as sites of corporate R&D. However, there is a notable difference between Seoul and Beijing, that in the case of Beijing, 95 percent of the articles come from oil firms (Sinopec, PetroChina, and CNOOC), while in Seoul, the most productive industry sector is electronics, having 42 percent of the articles. Andersson et al. (2014: 2969) claim that although Beijing has similar publication volumes as for example, London and Paris, it still occupies a peripheral position in the production of new scientific breakthroughs. Recently, by establishing technology parks, Beijing has made a strong effort to attract foreign high-technology firms and provide an innovative environment for domestic start-up firms (Zhou, 2005; Ramirez and Li, 2009; Zhang et al, 2011).

Outside the above regions, there are some less significant cities that are sites of corporate R&D; for example, Rio de Janeiro, Mumbai, and Melbourne, but together, they contribute only as many scientific articles as does the medium-sized city of Minneapolis.

## 5. Conclusion

In this paper, I conducted a bibliometric analysis to determine which cities were considered to be the international sites of corporate R&D, and how the publication performance of leading firms headquartered in these cities has changed between 1980 and 2014. Results show that cities can be organised into five large R&D zones worldwide: the West Coast of the United States, the Great Lakes Region, the East Coast of the United States, Western Europe, and East Asia.

*The West Coast of the United States*: The heart of this region is the San Francisco Bay Area (including San Jose and the cities of the Silicon Valley), which is the international centre of the information technology industry. The dominant city within this region is San Jose, having the largest number of scientific articles. Furthermore, it can be observed that many San Jose-based firms have started to publish only since the mid-1990s, and as such, San Jose is one of the most dynamically growing sites of corporate R&D worldwide. Seattle, Oxnard, San Diego, and Los Angeles are also significant sites of corporate R&D; however, their combined number of scientific articles is less than 50 percent of that of San Jose.

*The Great Lakes Region*: Chicago, Indianapolis, Detroit, Minneapolis, Cincinnati, and Midland are the most important sites of corporate R&D in this zone. It is a common characteristic of these cities that although large firms headquartered within them carry out significant R&D activities, and many of them operate in high-tech industries (e.g., chemicals, pharmaceuticals, and aerospace and defence), actually none of them belong to the fast-growing industries. Albeit the fact that cities in this zone are still important sites of corporate R&D, leading firms are publishing a decreasing number of articles year by year.

*The East Coast of the United States*: New York has the largest number of scientific articles published by firms, due to its globally outstanding pharmaceutical and information technology industry. New York is followed by some mid-sized R&D hubs, such as Bridgeport, Philadelphia, and Washington, DC. In this zone, the closest competitor to New York is Boston, the fastest-growing site of corporate R&D in the United States, especially in the fields of biotechnology and information technology.

*Western Europe*: On the basis of the number of scientific articles, Paris and London are the major sites of corporate R&D in Europe; however, with respect to the increase in the number of articles, cities in Southern Germany and Switzerland show the most rapid increase. Basel is a leading R&D hub of the international pharmaceutical industry. Furthermore, Copenhagen and Dublin also have a key role in the European corporate R&D. In the former, scientific articles primarily come from domestic biotechnology firms; in the latter, most of the articles are published by relocated American firms.

*East Asia*: The most dramatic changes have occurred in this region. Tokyo is the largest corporate R&D hub in the world: Tokyo-based firms published the largest combined number of articles in the period from 1980 to 2014; moreover, the city has had the largest annual output in terms of the number of scientific articles published since the mid-1980s. However, the volume of this annual output has been gradually decreasing; that is, year by year, Tokyo-based firms (and firms headquartered in Osaka and Nagoya) have been publishing fewer and fewer articles. In this region, Seoul and Beijing has become the most dynamically growing R&D hub, also reflected by the fact that firms headquartered in Seoul and Beijing have been publishing a rapidly growing number of scientific articles. The difference between Seoul and Beijing is that in the former, most articles are coming from high-tech firms, while in the latter, oil firms are publishing almost exclusively.

**Acknowledgement**

This paper is supported by the **János Bolyai Research Scholarship** of the Hungarian Academy of Sciences.**References**

Allen, R.C., 1983. Collective invention. *Journal of Economic Behavior and Organization*, 4(1): 1-24. DOI 10.1016/0167-2681(83)90023-9

Andersson, D.E., Gunessee, S., Matthiessen, C.W., Find, S., 2014. The geography of Chinese science. *Environment and Planning A*, 46(12): 2950-2971. DOI 10.1068/a130283p

Barrett, B., 2002. Defensive Use of publications in an intellectual property strategy. *Nature Biotechnology*, 20(2): 191-193. DOI 0.1038/nbt0202-191

Chang, K.-C., Chen, D.-Z., Huang, M.-H., 2012. The relationships between the patent performance and corporation performance. *Journal of Informetrics*, 6(1): 131–139. DOI 10.1016/j.joi.2011.09.001

Chang, Y.-W., 2014. Exploring scientific articles contributed by industries in Taiwan. *Scientometrics*, 99(2): 599–613. DOI 10.1007/s11192-013-1222-2

Cheng, S., 2014. *Financial Reforms and Developments in China*. World Scientific Publishing, Singapore.

Cohen, R. B., 1981. The new international division of labour, multinational corporations and urban hierarchy. In: Dear, M., Scott. A. (eds), *Urbanization and Urban Planning in Capitalist Societies*. Methuen, London-New York: 287-316.

Csomós G., Tóth G., 2016a. Exploring the position of cities in global corporate research and development: A bibliometric analysis by two different geographical approaches. *Journal of Informetrics*, 10(2): 516-532. DOI 10.1016/j.joi.2016.02.004

Csomós G., Tóth G., 2016b. Modelling the shifting command and control function of cities through a gravity model based bidimensional regression analysis. *Environment and Planning A*, 48(4): 613-615. DOI 10.1177/0308518X15621632

Dernis H., Dosso M., Hervás F., Millot V., Squicciarini M., Vezzani A., 2015. *World Corporate Top R&D Investors: Innovation and IP bundles. A JRC and OECD common report*. Publications Office of the European Union, Luxembourg.

Dicken, P., 2007. *Global Shift: Mapping the Changing Contours of the World Economy, 5th Edition*. Sage Publication, London.

Feng, F., Zhang, L., Du, Y., Wang, W., 2015. Visualization and quantitative study in bibliographic databases: A case in the field of university-industry cooperation. *Journal of Informetrics*, 9(1): 118–134. DOI 10.1016/j.joi.2014.11.009

Financial Times, 2012. Alcatel at fault for problems, says chief. Last accessed: July 7, 2016 Online: http://www.ft.com/cms/s/0/46a5637a-db18-11e1-8074-00144feab49a.html#axzz4DilrkW22 (accessed 7 July 2016)

Fröbel, F., Heinrichs, J., Kreye, O., 1980. *The New International Division of Labour: Structural Unemployment in Industrialised Countries and Industrialisation in Developing Countries*. Cambridge University Press, Cambridge.

Furukawa, R., Goto, A., 2006. Core scientists and innovations in Japanese electronics companies. *Scientometrics*, 68(2): 227–240. DOI 10.1007/s11192-006-0109-x14